%% file: sample-sigconf.tex
\documentclass[sigconf]{acmart}

\usepackage{booktabs} 
\usepackage{hyperref}

\usepackage{graphicx}
\usepackage{blkarray}
\usepackage{subfigure}

\setcopyright{none}

\begin{document}
\title{Classical Music Clustering Based on Acoustic Features}

\author{Xindi Wang}
\affiliation{%
  \institution{Northeastern University}
  \streetaddress{360 Huntington Ave.}
  \city{Boston} 
  \state{Massachusetts} 
  \postcode{02115}
}
\email{wang.xind@husky.neu.edu}

\author{Syed Arefinul Haque}
\affiliation{%
  \institution{Northeastern University}
  \streetaddress{360 Huntington Ave.}
  \city{Boston} 
  \state{Massachusetts} 
  \postcode{02115}
}
\email{haque.s@husky.neu.edu}

\renewcommand{\shortauthors}{Syed Arefinul Haque, Xindi Wang}

\begin{abstract}
In this paper we cluster 330 classical music pieces collected from MusicNet database based on their musical note sequence. We use shingling and chord trajectory matrices to create signature for each music piece and performed spectral clustering to find the clusters. Based on different resolution, the output clusters distinctively indicate composition from different classical music era and different composing style of the musicians.
\end{abstract}

%
%





\maketitle

\input{samplebody-conf}

\bibliographystyle{ACM-Reference-Format}
\bibliography{sigproc} 

\end{document}

%% file: samplebody-conf.tex
\section{Introduction}

For centuries, people have been intrigued by the depth and richness of classical music. This leads to questions like what characterize Bach's music and what makes them so different from others compositions. In this project, we build a clustering algorithm for classical music pieces from the music itself using MusicNet dataset~\cite{thickstun2017learning}. This clustering algorithm would help music recommendation, understand patterns in music and relationships between music pieces.

\subsection{Motivation}
\subsubsection{Genres we get from human cooperative classification are fuzzy}

Categorizing musics based on genre is a difficult problem. Currently, people rely on methods involving humans, such as social tags and collaborative filtering. However, social tags are subjective decisions which could be influenced by cultural and perceptual differences~\cite{Barreira2011}. Collaborative filtering~\cite{Cano2006} might not work well for music pieces that have not appeared in anyone's playlist or review.

\subsubsection{Acoustic feature to understand music compositions}

One way to address this problem is to rely on acoustic features. This prevents the incoherency related to human labeling and this is the only kind of feature that is presented in any music. Also, during the process, we would have the opportunity to gain patterns in music compositions.

\subsection{Contribution}
\begin{itemize}
    \item Built a system that could take any given set of music with their note sequence, find clusters based on two extracted features.
    
    \item Proposed two ways of extracting features from music's note sequence, shingles and chord trajectory matrix. Among them, chord trajectory matrix is novel and offers us a new way to see music structure.
    
    \item Found clusters that are related to composers, and could reflect music era.
\end{itemize}

\section{Related Work}

In music feature extraction, spectrogram~\cite{isaacson2005you} is a very widely used feature. A spectrogram is a visual representation of the spectrum of frequencies in a sound or other signal as they vary with time or some other variable. The calculation of spectrogram is related to Fourier transformation. However, spectrogram has limited potential to help us understand composer's composing styles and patterns. Spectrogram only takes care of spectrum of a given music piece, not focusing on how music progresses, how different parts of music cooperate together, etc., and is highly related to instruments. 

In~\cite{Vitanyi2004}, an information theoretic approach was applied to find the similarity between musics. They used a dataset of MIDI files of 60 classical music pieces, 12 jazz pieces and 12 rock pieces and clustered those musics based on normalized compression distance (NCD). They mentioned that an ideal candidate for the information theoretic distance would have been Kolmogorov complexity which is not computable in practice. Instead they rely on the compressibility of music as a proxy of information theoretic distance. Genres they found using this method seems to conform to the specific artist and pre-labeled genres. The quartet method they used produces slightly different similarity score even for the same pair of music because of the randomness in the similarity calculation process.

Since the purpose of our clustering is not only finding similar pieces, but also discovering composing style for different composers, we choose to utilize information directly related to composition - the music note sequence, and propose method that could compare different note sequences.

\section{Background}

\subsection{Basic Music Concept}
    \subsubsection{Measure}
    In musical notation, a measure (or bar) is a segment of time corresponding to a specific number of beats in which each beat is represented by a particular note value and the boundaries of the bar are indicated by vertical bar lines~\cite{read1972music}.
    
    \subsubsection{Beats}
    In music theory, the beat is the basic unit of time, the pulse (regularly repeating event), of the mensural level~\cite{read1972music}. In MusicNet data, each note has a record of measure and beat. Here the beat denotes the time when this note happens in this specific beat, where each measure has time interval "1". For example, in the score below, the first note has beat 0, the second note has beat 1/4 (since time signature is 4/4, so each quarter note has a length of 0.25).
    
    \begin{figure}[!h]
    \includegraphics[width=2.5in]{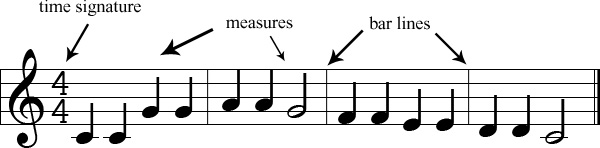}
    \caption{Concepts of classical music}
    \end{figure}

\subsection{Music Coding}
    \subsubsection{Instrument Code}
    MusicNet records the instrument of each note as well. The coding is based on the MIDI instrument code and the decoding information for different instruments can be found in~\cite{midi_instruments}.
    
    \subsubsection{MIDI Codes for notes}
    To record the note value, MusicNet use MIDI encoding for notes, which ranges from 0-127. It expands 11 octaves for 12 pitches (C, C\#, etc.). Information for decoding the midi code for notes can be found in ~\cite{midi_sequences}.

    \subsubsection{Interval Tree}
    \label{sec:intervaltree}
    MusicNet data use Interval Tree~\cite{de2000computational} to store information of notes. An interval tree is a tree data structure to hold intervals. Specifically, it allows one to efficiently find all intervals that overlap with any given interval or point. It is often used for windowing queries.
    
\begin{figure*}[t]
    \includegraphics[width=6in]{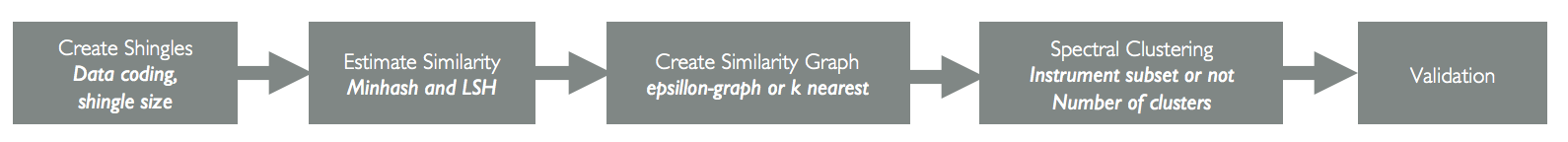}
\caption{Flow chart for shingles}
\label{fig:flowchart(shingle)}
\end{figure*}

\begin{figure*}[t]
    \includegraphics[width=6in]{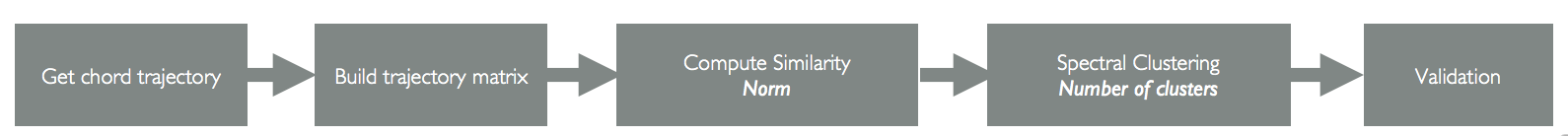}
    \caption{Flow chart for chord trajectory matrix}
    \label{fig:flowchart(chord)}
\end{figure*}

\section{Proposed Approach}

The question we want to solve is, given a set of music, how we can get individual signature of each music by utilizing only musical notation features, and then how we can find clusters and validate them.

\subsection{Feature Extraction}

\subsubsection{Shingles}
A natural way of extracting musical features is by regarding each music piece as a document, where the "alphabets" are the notes. Therefore, we could use shingling method to build the signature-music matrix and use MinHash and LSH to estimate similarity. Through MinHash and LSH we can query pieces having similarity larger than a threshold we want or get top $k$ similar pieces for each piece. Therefore, we could produce a $\epsilon$-affinity matrix or a $k$-nearest neighbor affinity matrix.

\subsubsection{Chord Trajectory Matrix}
Another way of extracting music features is through building a Chord Trajectory Matrix. In music, chord transition is very important and varies based on different composers, instruments and themes. Since we have 128 MIDI notes (coded 0-127) and pause (we coded it as 128), we would have a $129 \times 129$ matrix that could represent one music piece. After building Chord Trajectory Matrices, we could use different norms to calculate similarity between them and obtain similarity score for each pair of music pieces.

\subsection{Spectral Clustering}
Both shingles and chord trajectory matrix would produce an affinity matrix for music pieces, where an entry close to 0 means the pair is dissimilar and close to 1 means they are similar. Spectral clustering~\cite{shi2000normalized} is a clustering method focusing on the connectivity of data, which directly utilize similarity between data points. To perform spectral clustering we need constructs the Laplacian matrix from the affinity matrix and use the eigenvectors to obtain balanced cut of the graph.

\subsection{Validation}
Validation of clusters is a difficult problem. In this project, we use both internal and external validation. For internal validation, we use silhouette coefficient, which compare the distances within cluster and between clusters. For external validation, we utilize the meta-data we have, which includes composers, movements and ensembles. We put all the meta-data together and form a "document" for each cluster. Then we calculate TF-IDF~\cite{leskovec2014mining} for each term to help us understand the cluster as well as validating them. For example, to get an idea of which cluster is mostly about which musician we get the composer metadata for each music piece in the cluster and find the TF-IDF value for all possible composers in that cluster. Then we choose the top $k$ keywords(composer) for the cluster to be its sense making keyword.

\subsection{Flow Chart}
The flow chart of our approach is shown in Fig.~\ref{fig:flowchart(shingle)} and Fig.~\ref{fig:flowchart(chord)}

\section{Experiments}

\subsection{Data and Data Processing}
The dataset we use is MusicNet~\cite{thickstun2017learning}, which is a collection of 330 free-licensed classical music recordings, in total of 34 hours of chamber music performances. It consists of almost 1.3 million labels indicating the precise time of each note every recording, the instrument that plays each note and the notes position in the metrical structure of the composition. Besides important Acoustic Features, MusicNet also includes composer and title. These meta-data label for clustering are used during validation of results.

The data structure of MusicNet is interval tree (See \hyperref[sec:intervaltree]{3.2.3}), which allows users to query what notes are playing at a given time or between a time interval. However, querying by time would produce duplicate note record if the length of the note is longer than the query interval and this would influence the quality of shingles and chord trajectory matrix. Therefore we develop an algorithm to extract note sequence using the information of start-time, end-time, measure, beat for each individual note. 

The key of this algorithm is to add pauses between notes because we would lost information of pauses if we use all the labels in the interval tree directly. Adding pauses are done in two parts: adding pauses between notes and adding pauses at beginning of a measure and at end of a measure. For adding pauses between notes, we compare the start time and end time for sequential notes and if the difference is larger than a threshold, we add a pause between them. For adding pauses at the beginning and the end, we obtain the start time and end time for each measure across all instruments, naming global start time and global end time. Then for each measure for each instrument, we compare its start and end time with the global start and end time. If the difference is larger than a threshold, we add a pause.

The limitation of this process is that, we only add one pause regardless of how long the pause is. This may influence the result of the shingles and the chord trajectory matrix.

Another problem in the dataset is that the note are only distinguished by instrument. However, when we have two violins in one piece, we can not distinguish them. This problem could potentially be solved using factorization. For simplicity, we ignore this issue in the project. Therefore the note sequence we get is merged as one instrument for each unique type of instrument.

One final problem is how to create shingles out of note sequence with notes being played together. There are multiple ways to deal with this problem. For the sake of simplicity we take the numerically highest note if there are multiple notes happening at the same time. This is based on the assumption that notes appearing together usually constitutes a harmonic chord, and the highest note usually could represent the dominant note in this chord.

\subsubsection{Coding Variation}
The original data coding for note is the 128 MIDI note coding. We have different variation of coding to capture different aspect of the note.

\paragraph{Pitch Coding}
Ignoring the octaves, we could convert the 128 MIDI coding to 12 pitch coding. The 12 pitches are C, C\#, D, D\#, E, F, F\#, G, G\#, A, A\#, B. In this way, we only focus on what chord is now presenting.

\paragraph{Relative Coding}
Recoding the first note of the piece as 0, we could recode every note in this piece based on this first note baseline. In this way, we would have similarity 1 if two pieces are just a simple modulation shift.

\paragraph{Change Coding}
We could calculate the change between two sequential notes by simply subtracting the MIDI code of the later note from the MIDI code of previous note. Then we can use the sequence of differences to represent the music. In this way, we are focusing on the change of chord of the music piece.

\subsection{Results}
\subsubsection{Shingles}
We did shingling on different shingle size ($k \in [2,7]$) and different data coding (default MIDI coding, pitch coding, relative coding, change coding). To get affinity matrix, we use MinHashing and LSH to efficiently calculate similarity and obtain two types of graph, $\epsilon$-graph and $k$-nearest neighbor graph. For $\epsilon$-graph, we use different threshold for different size of shingle ($0.1$ for $k = 2$, $0.05$ for $k = 3$, $0.01$ for $k = 4$ and $k = 5$, $0.005$ for $k = 6$ and $k = 7$). For $k$-nearest neighbor graph, we choose the number of neighbors as $10$. When doing spectral clustering, we could use the whole corpus or select specific instrument we are interested in - we select piano (MIDI code 1), violin (MIDI code 41), viola (MIDI code 42) and cello (MIDI code 43) in our experiments.

\begin{figure}[!h]
    \includegraphics[width=2.5in]{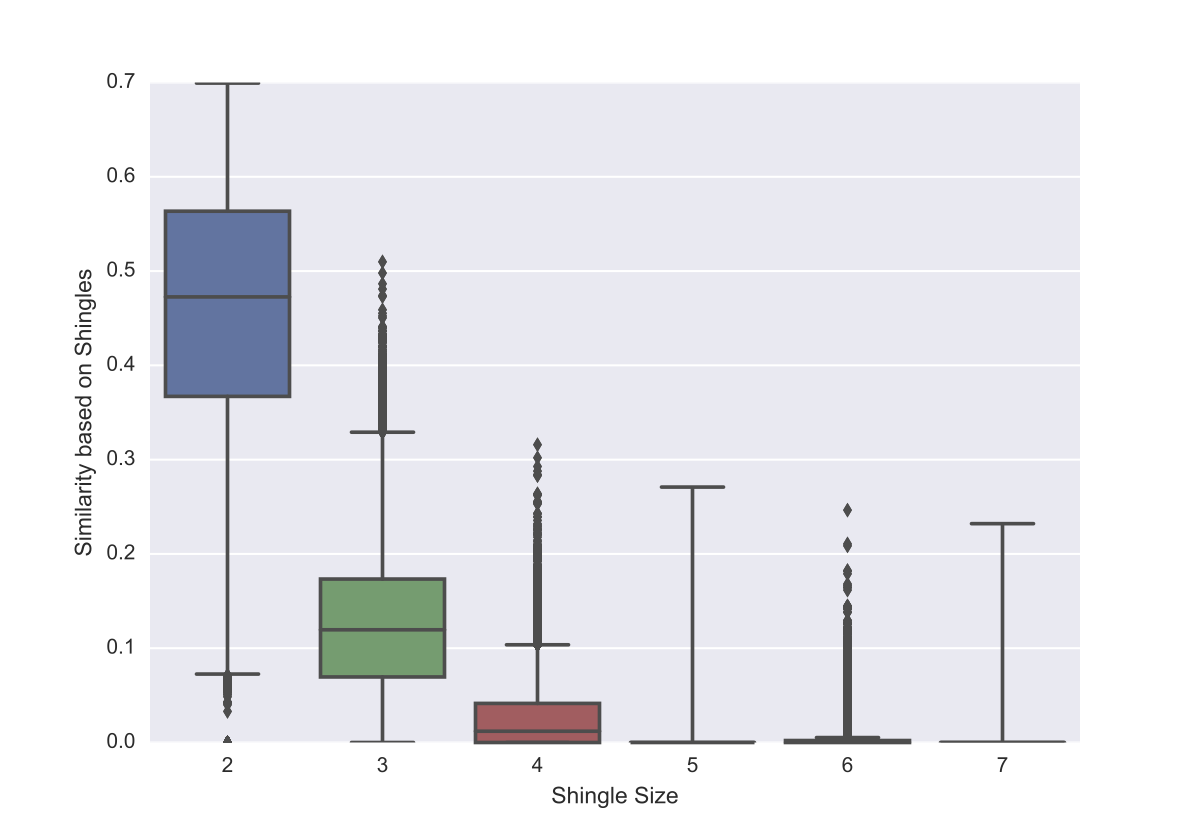}
    \caption{Similarity distribution for different shingle size}
    \label{fig:sim}
\end{figure}

The obtained similarity distribution for different shingle size is shown in Fig.~\ref{fig:sim}. We could see that the similarity score goes down very quickly when we have a larger shingle size. For $k = 2$, the average similarity score is about 0.5, but for $k=7$, most of the pieces has similarity close to 0.

\begin{figure}[!htb]
\subfigure{\includegraphics[width=2.5in]{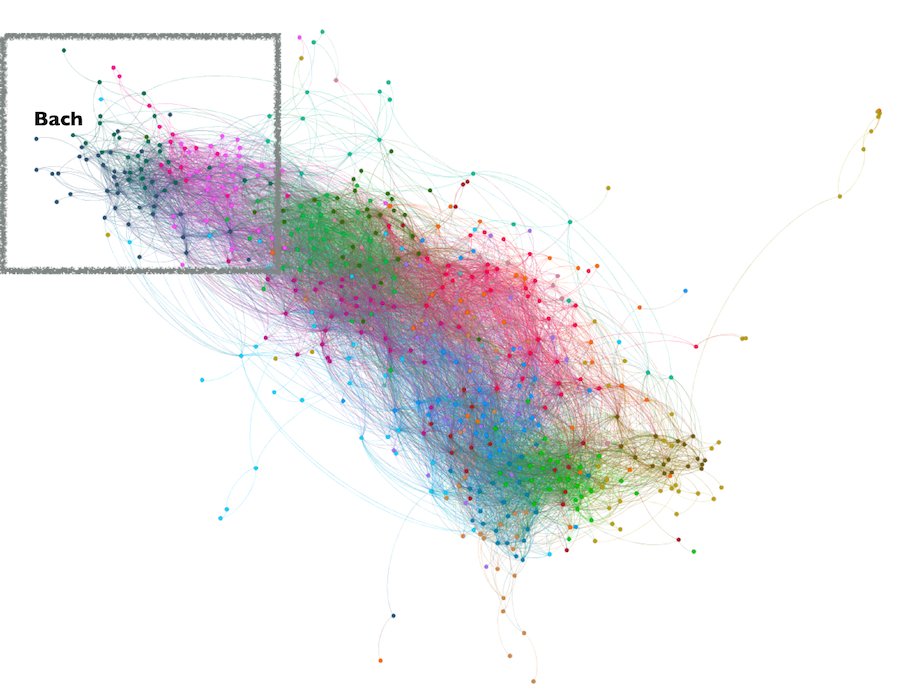}}
\subfigure{\includegraphics[width=2.5in]{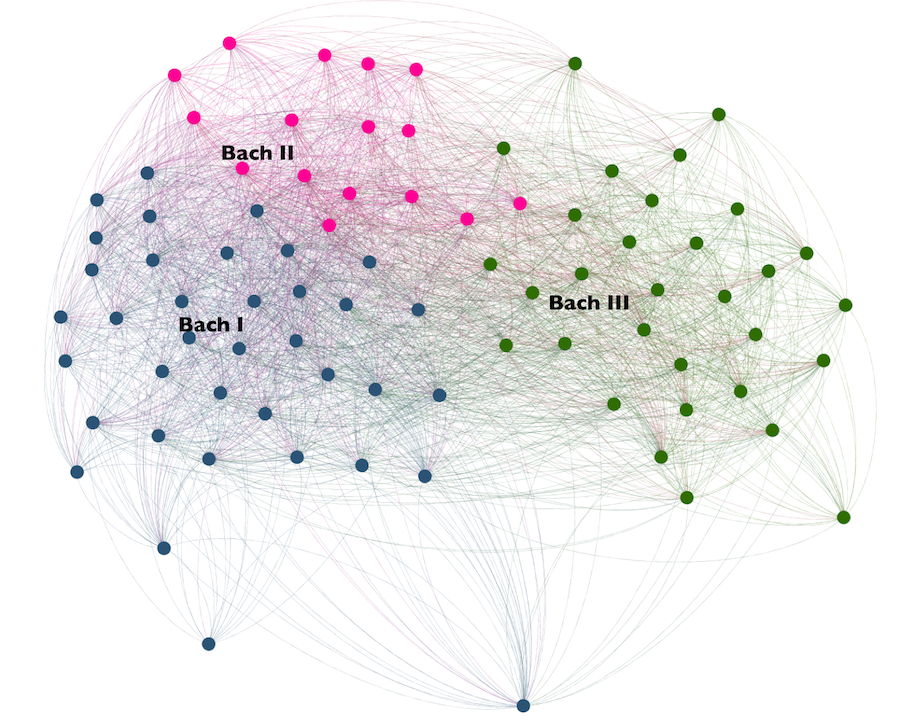}}
\caption{Cluster visualization for shingle method (22 clusters)}
\label{fig:cluster shingle}
\end{figure}

Here is one example clustering result on shingle size 4, using pitch coding and $\epsilon$-graph, clustered on the whole corpus and setting number of clusters to be 22 (Fig.~\ref{fig:cluster shingle}). Using TF-IDF~\cite{leskovec2014mining} to find the characteristic keywords of each cluster, we could see that on the left of the graph, there is a big cluster of Bach music. When we zoom into this cluster, there are actually three clusters all containing Bach musics. The keyword of the first one is "fugue-Bach", the second one is "cello-suite-Bach" and the third one is "harpsichord-Bach". This cluster separation corresponds to different types of Bach music we have in our corpus. Therefore, we could conclude that clusters we found could reflect certain work of a composer.

\begin{figure}[!h]
    \includegraphics[width=2.5in]{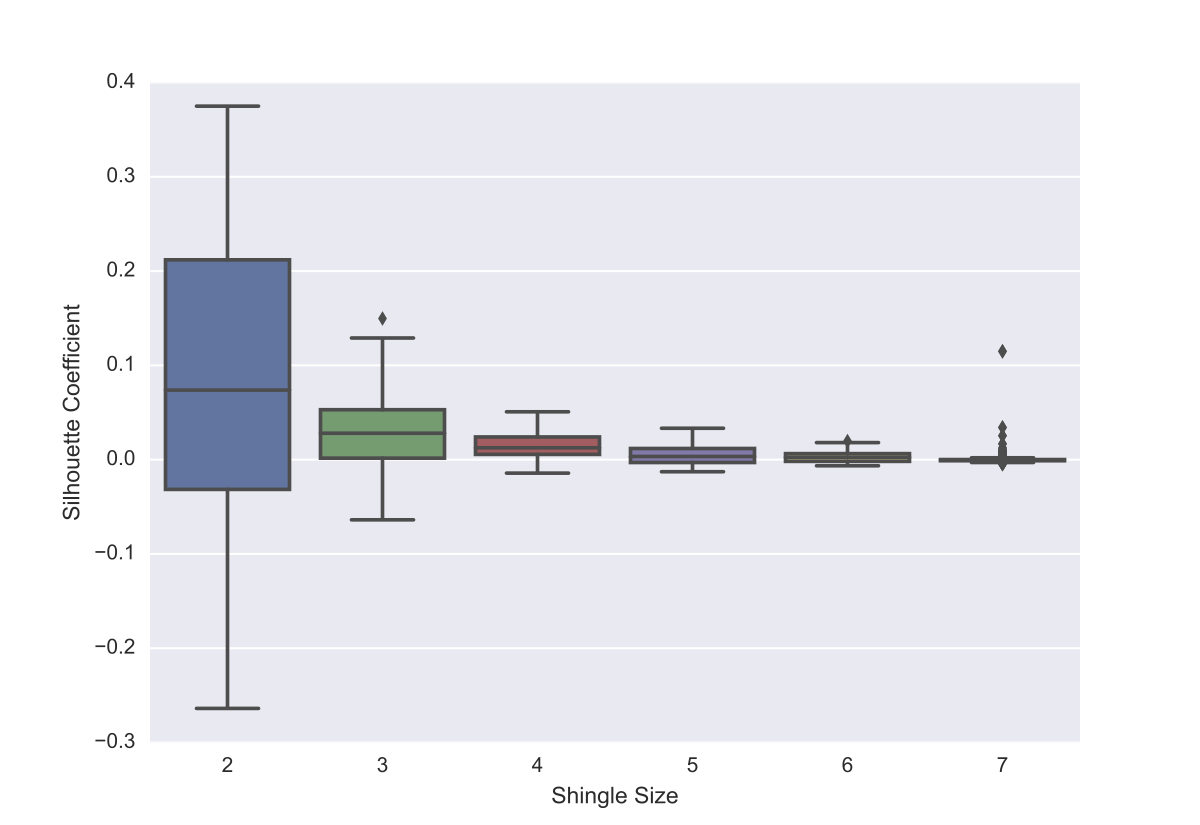}
\caption{Silhouette Coefficient for different shingle size (10 clusters)}
\label{fig:SC_shingle}
\end{figure}

\begin{figure}[!h]
    \includegraphics[width=2.5in]{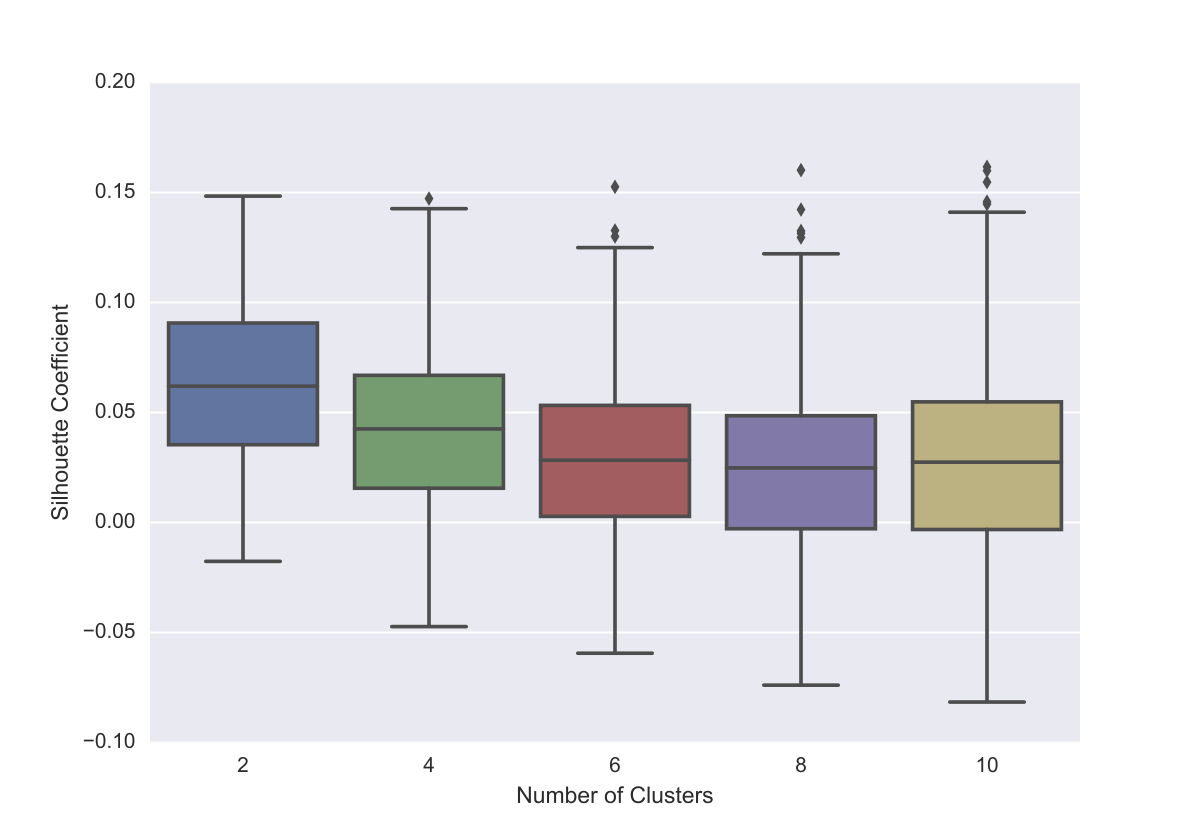}
    \caption{Silhouette Coefficient for different number of clusters (shingle size 3)}
    \label{fig:SC_cluster}
\end{figure}

To further validate the performance of clustering, we use the internal method for cluster validation - silhouette coefficient~\cite{rousseeuw1987silhouettes}. Silhouette coefficient compare the within cluster distance and between cluster distance. The closer to 1, the better the performance. Fig.~\ref{fig:SC_cluster} shows the silhouette coefficient distribution for different shingle size fixing the number of clusters to be 10 (here we all use pitch coding). We could see that as the shingle size increases, the silhouette coefficient decreases. Fig.~\ref{fig:SC_shingle} shows the silhouette coefficient distribution for different number of clusters setting shingle size to be 3. With more clusters, the silhouette coefficient score decreases. However we could see that the silhouette coefficient score is low in general, far from the ideal value of 1. This may be because the similarity score between pieces are very low in general (shown in Fig.~\ref{fig:sim}), therefore the within cluster distance and between cluster distance are not that different from each other.

We also conduct K-medoids clustering on shingles. When we use the Manhattan distance, the output is always a huge cluster even when we increase the number of cluster. When we use Jaccard distance, we have similar result as spectral clustering and the silhouette coefficient remains to be low. For $k=3$ and pitch coding, the average silhouette coefficients are $0.0176$,$0.0126$, $0.0063$, for the number of clusters $2$, $5$ and $10$ respectively. From silhouette coefficient point of view, spectral clustering is better than K-medoids.

\subsubsection{Chord Trajectory Matrix}

The chord trajectory matrix is another way to create a signature out of each music piece. In this matrix we create a relationship between each MIDI code (ranging from 0-127) and pauses (considered as the 128th note) based on which note follows which note. In other words we can think of the sequence of musical notes as a directed path and chord trajectory matrix is the adjacency matrix of that path. To clarify the concept lets imagine we have the following sequence of musical notes: $\{A, C, C, A, C, pause, C, C, pause, pause\}$. Then without considering the MIDI encoding we can think of the sequence as a set of the edges $\{(A,C),(C,C),(C,A),...,(pause,pause)\}$ and convert them to the chord trajectory matrix,$C_{3\times 3}$ which looks like the following:

\begin{equation*}
    \mathbf{C} = 
    \begin{blockarray}{cccc}
    A & C & pause \\
        \begin{block}{(ccc)c}
          0 & 2 & 0 & A \\
          1 & 2 & 2 & C \\
          0 & 1 & 1 & pause \\
        \end{block}
    \end{blockarray}
\end{equation*}

In the above chord trajectory matrix, we make an increment to an entry of C, $c_{ii}$ by 1 when a note $i$ is followed by the note itself and we make an increment to the entry of C, $c_{ij}$ by 1 when note $i$ is followed by note $j$. But there are some cases where a collection of notes are played together and followed by another note/s. For example lets imagine the notes' sequence is  $\{\{A,B\},\{B,C\},D\}$. In that case the edges would be cross product of each pair of sets in the sequence and the edges for the chord trajectory matrix would look like $\{(A,B),(A,C),(B,B),(B,C),(B,D),(C,D)\}$. \\

\begin{figure*}[h]
    \includegraphics[width=0.56\textwidth]{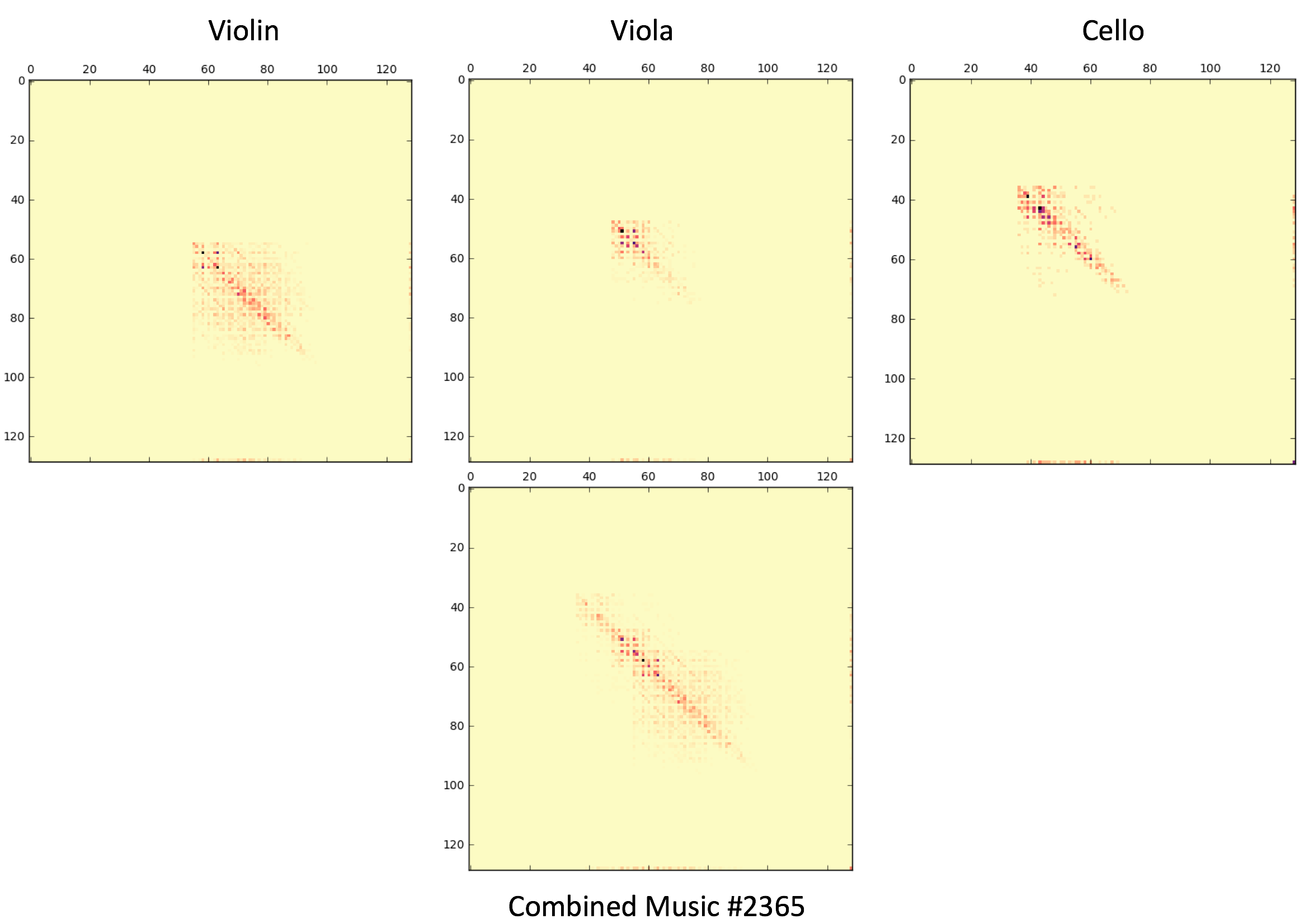}
\caption{Combining different instruments in an ensemble into single chord trajectory matrix}
\label{fig:combining_instruments}
\end{figure*}

\textbf{Combining chord trajectory matrices in an ensemble piece:} If a music piece is an ensemble of different instruments then we create the chord trajectory matrix of each instrument separately and finally add them together. All of the chord trajectory matrix is $128 \times 128$. As a result, simple addition between the matrices are well defined. In Fig.~\ref{fig:combining_instruments} we show one such example where we create a chord trajectory matrix of music no. 2365 from the MusicNet dataset which is titled String Quartet No 12 in E-flat, composed by Beethoven.

\begin{figure*}[h]
    \includegraphics[width=0.56\textwidth]{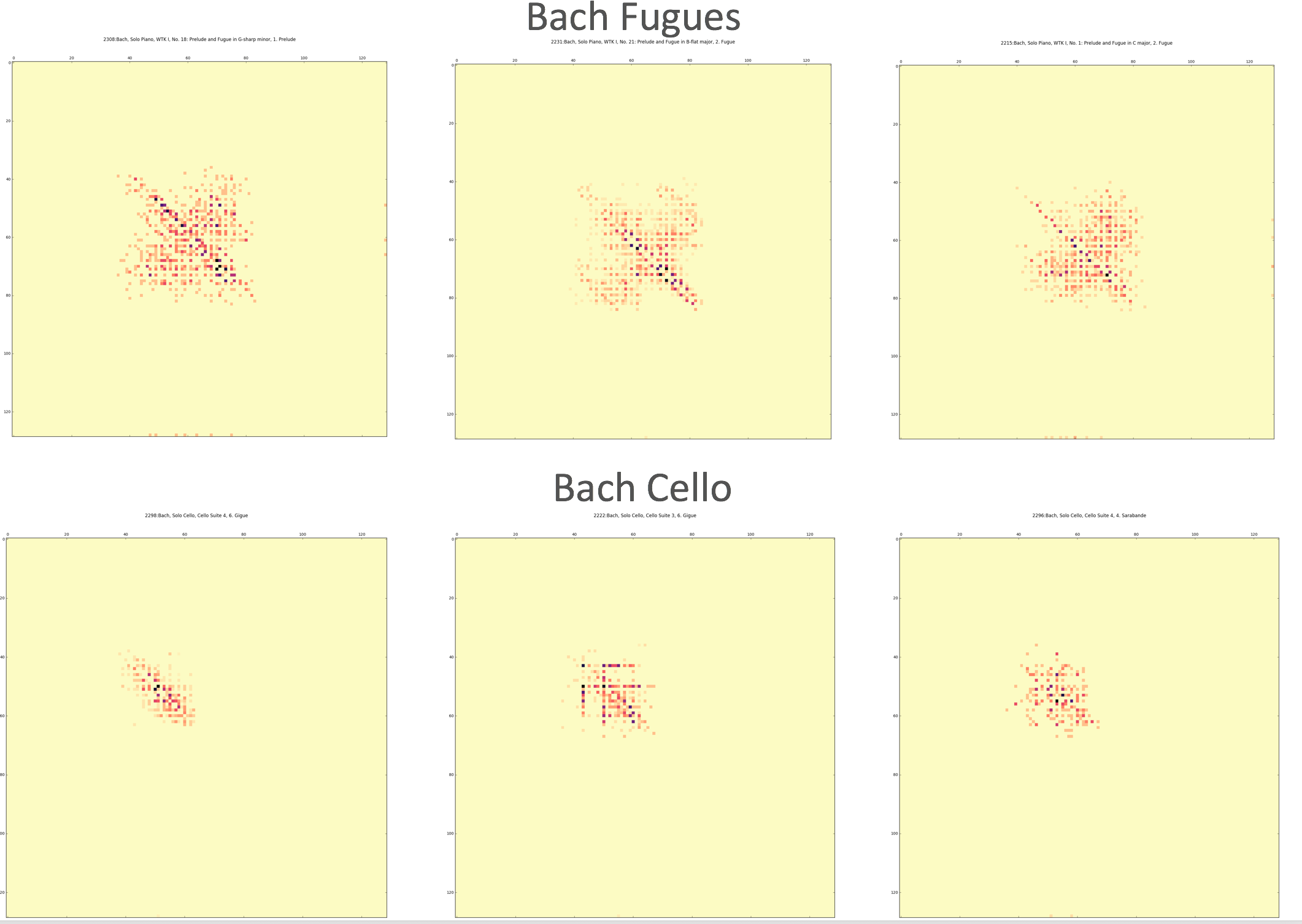}
\caption{Patterns and similarity in Bach music}
\label{fig:bach}
\end{figure*}

\textbf{Visual similarity in chord trajectory matrix:} By inspecting the chord trajectory matrix, we can see some interesting visual similarity between the music pieces. We know that Bach is celebrated for his unique style of composing fugue and cello compositions. By inspecting the visual representation of the chord trajectory matrices In Fig.~\ref{fig:bach} we can see that Bach's Fugues have an unique shape which looks like Phoenix bird and his cello compositions almost has a beetle like shape. The most intriguing outcome from the visual inspection is that each of the Bach fugue has this unique phoenix bird shape.

\textbf{Finding difference between two chord trajectory matrices}: To find the distance between two chord trajectory matrices $C_1, C_2$, at first we create a difference matrix, $D = C_1 - C_2$. Then we calculate the norm of $D$. We have tried two different norms. The first norm is $||D||_F = \sqrt{\sum_i{\sum_j{d_{ij}^2}}}$, which is known as the Frobenius norm and second norm is , $||D||_F = {\sum_i{\sum_j{|d_{ij}|}}}$ which we call as absolute value norm. Upon inspection we have found that absolute value norm penalizes small differences in the entries of two matrices which is not desired for our distance calculation. So we finally used Frobenius norm to calculate distances between two matrices.

\textbf{Creating affinity matrix}: After finding pairwise distances between all the 330 music we create an affinity matrix. But before that we normalize all the distances by doing a feature scaling using the following formula: $x' = \frac{x-min(X)}{max(X) - min(X)}$. After normalizing we convert the distance into affinity by subtracting the distance from 1.

\textbf{Spectral Clustering:} Finally from the affinity matrix we perform spectral clustering in the some method as described above. To validate the clustering we calculated the silhouette co-efficient of the whole cluster. The co-efficient is very low for all the cases. We tried different number of clusters. the average silhouette coefficients are $0.232$, $0.106$ and $0.021$ respectively when the number of clusters are 2, 5 and 24.

\begin{figure*}[h]
    \includegraphics[width=0.5\textwidth]{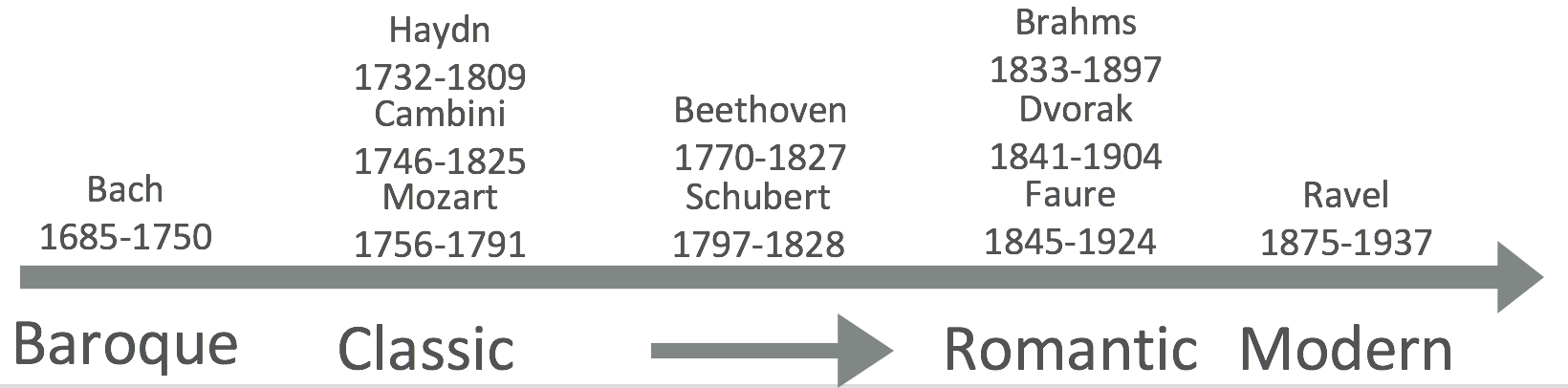}
\caption{Era of classical music}
\label{fig:classical_music_era}
\end{figure*}

\begin{figure}[h]
    \includegraphics[width=2.5in]{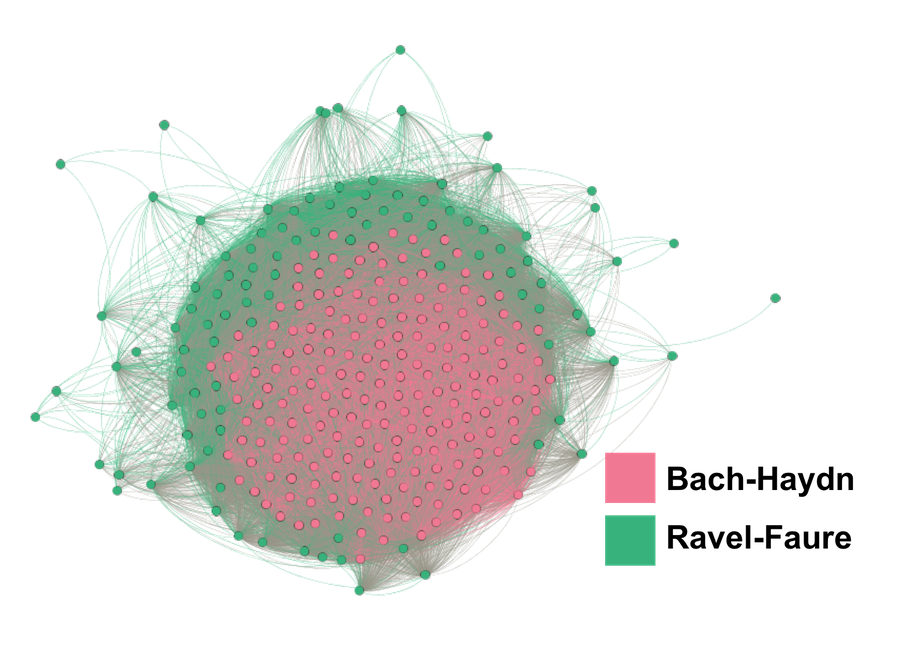}
\caption{Cluster visualization for chord trajectory matrix (2 clusters)}
\label{fig:cluster2}
\end{figure}

\begin{figure}[h]
    \includegraphics[width=2.5in]{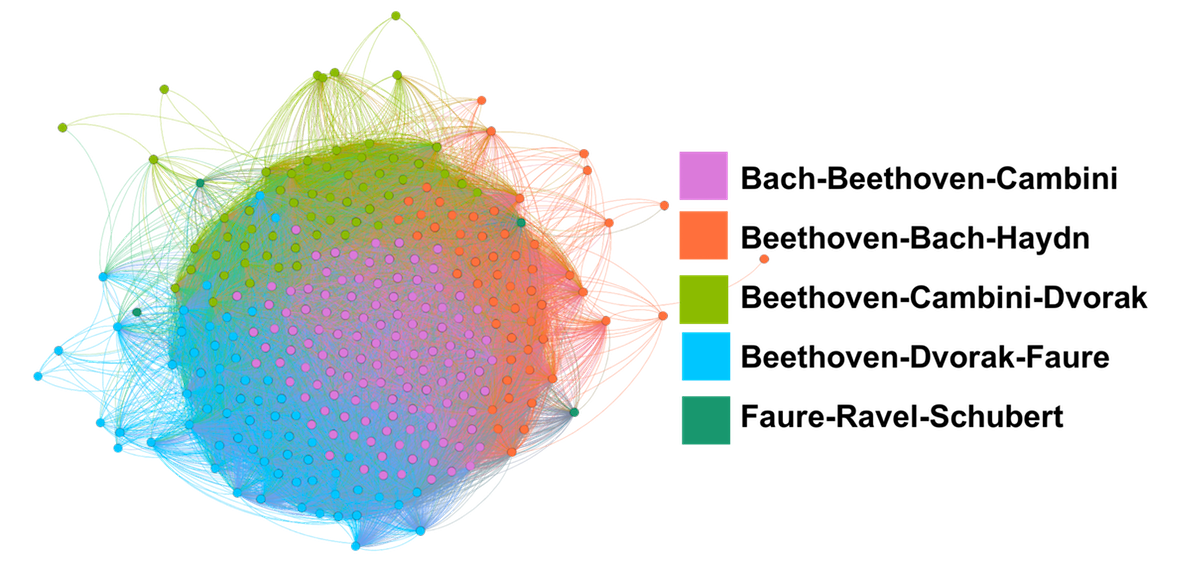}
    \caption{Cluster visualization for chord trajectory matrix (5 clusters)}
    \label{fig:cluster5}
\end{figure}

\textbf{Sense Making of the Clusters:} When we find two clusters in all the 330 music pieces (Fig.~\ref{fig:cluster2}), the first cluster's keyword is "Bach-Haydn" and the other cluster's keyword is "Ravel-Faure". Which makes sense based on the division of epochs in classical music. Based on the era of different classical musicians in Fig.~\ref{fig:classical_music_era} we can say that these two clusters indicate the two extremes of the timeline.\\ Again, when we break them into five clusters (Fig.~\ref{fig:cluster5}) we can see that each of the clusters represent a sequential mix of different musical era demonstrated in Fig.~\ref{fig:classical_music_era}.

Most interestingly, we see all the Bach musics are contained in a single large cluster along with other pieces while we keep increasing the number of clusters; although we saw the visual dissimilarity between Bach's fugues and cello compositions. When the number of clusters is increased into 24, suddenly Bach musics are nicely separated into two clusters, where one cluster is all the fugues along with some compositions of Beethoven and in the other cluster we find all the cello compositions by Bach along with a single piano composition by him.

\section{Conclusions and Future Work}

\subsection{Conclusions}
In this project, we develop a process for music clustering using two set of feature extraction methods - shingles and chord trajectory matrix. We find that the clusters for both features are related to composers style of composing music. For chord trajectory matrix approach, the clusters we found also relate to different eras of music. From chord trajectory matrix itself, we also discover certain patterns for musical ensembles and composers.

\subsection{Limitations and Future Work}
\begin{itemize}
    \item \textbf{Improvements on shingles} From our investigation, clustering results for shingles are not ideal based on our meta-data. The clusters are more mixed than the chord trajectory approach. This could relate to the choice of shingle size and data coding. 
    
    \item \textbf{Improvements on chord trajectory matrix} The chord trajectory matrix we proposed are very basic. We omit the information of instrument and ensembles. Also the norm we used for similarity calculation are basic Frobenius norm and Absolute value norm. In our future work we would like to put weights on  instrument information while constructing chord trajectory matrix and we will also explore different norms to find the similarity score between matrices.
    
    \item \textbf{Investigation on silhouette coefficient} We use silhouette coefficient as the internal validation for clusters. However the result is not ideal. We would like to investigate more on whether silhouette coefficient is a good way for our clusters and method and whether there are better metric for cluster validation. 
    
    \item \textbf{Analysis on harmony} In our proposed feature extraction methods (shingles and chord trajectory matrix), we emphasize on how music progress, i.e notes appearing sequentially. However, another very important aspect of music is about what notes are appearing together to produce harmony. We would like to include this feature into our future analysis and clustering.
    
    \item \textbf{Temporal information} The features we use all lose the temporal information. We would like to include the time information in our features by using methods such as temporal motifs~\cite{thickstun2017learning}.
    
    \item \textbf{Data limit} We only have 330 pieces of classical music data, and the data is biased. We have about 40\% percent of Beethoven music and the remaining 60\% percent belongs to 9 different composers. This leads to the fact that Beethoven appears in almost every cluster we have. We would like to experiment our method on a larger corpus.

\end{itemize}

%% file: sample-sigconf.bbl

\begin{thebibliography}{00}


\ifx \showCODEN    \undefined \def \showCODEN     #1{\unskip}     \fi
\ifx \showDOI      \undefined \def \showDOI       #1{#1}\fi
\ifx \showISBNx    \undefined \def \showISBNx     #1{\unskip}     \fi
\ifx \showISBNxiii \undefined \def \showISBNxiii  #1{\unskip}     \fi
\ifx \showISSN     \undefined \def \showISSN      #1{\unskip}     \fi
\ifx \showLCCN     \undefined \def \showLCCN      #1{\unskip}     \fi
\ifx \shownote     \undefined \def \shownote      #1{#1}          \fi
\ifx \showarticletitle \undefined \def \showarticletitle #1{#1}   \fi
\ifx \showURL      \undefined \def \showURL       {\relax}        \fi
\providecommand\bibfield[2]{#2}
\providecommand\bibinfo[2]{#2}
\providecommand\natexlab[1]{#1}
\providecommand\showeprint[2][]{arXiv:#2}

\bibitem[\protect\citeauthoryear{??}{mid}{2016a}]%
        {midi_instruments}
 \bibinfo{year}{Accessed April 25, 2016}\natexlab{a}.
\newblock \bibinfo{booktitle}{{\em General MIDI instrument codes}}.
\newblock
\showURL{%
\url{http://www.ccarh.org/courses/253/handout/gminstruments}}


\bibitem[\protect\citeauthoryear{??}{mid}{2016b}]%
        {midi_sequences}
 \bibinfo{year}{Accessed April 25, 2016}\natexlab{b}.
\newblock \bibinfo{booktitle}{{\em MIDI Note Numbers for Different Octaves}}.
\newblock
\showURL{%
\url{http://www.electronics.dit.ie/staff/tscarff/Music_technology/midi/midi_note_numbers_for_octaves.htm}}


\bibitem[\protect\citeauthoryear{Barreira, Cavaco, and Ferreira}{Barreira
  et~al\mbox{.}}{2011}]%
        {Barreira2011}
\bibfield{author}{\bibinfo{person}{Luis Barreira}, \bibinfo{person}{Sofia
  Cavaco}, {and} \bibinfo{person}{Joaquim Ferreira}.}
  \bibinfo{year}{2011}\natexlab{}.
\newblock \showarticletitle{{Unsupervised Music Genre Classification with a
  Model-Based Approach}}.
\newblock  (\bibinfo{year}{2011}), \bibinfo{pages}{268--281}.
\newblock


\bibitem[\protect\citeauthoryear{Cano, Celma, Koppenberger, and Buld}{Cano
  et~al\mbox{.}}{2006}]%
        {Cano2006}
\bibfield{author}{\bibinfo{person}{Pedro Cano}, \bibinfo{person}{Oscar Celma},
  \bibinfo{person}{Markus Koppenberger}, {and} \bibinfo{person}{Javier~M.
  Buld}.} \bibinfo{year}{2006}\natexlab{}.
\newblock \showarticletitle{{Topology of music recommendation networks}}.
\newblock \bibinfo{journal}{{\em Chaos\/}} \bibinfo{volume}{16},
  \bibinfo{number}{1} (\bibinfo{year}{2006}).
\newblock
\showISBNx{1054-1500 (Print)}
\showISSN{10541500}
\showDOI{%
\url{https://doi.org/10.1063/1.2137622}}
\showeprint[arxiv]{arXiv:physics/0512266v1}


\bibitem[\protect\citeauthoryear{De~Berg, Van~Kreveld, Overmars, and
  Schwarzkopf}{De~Berg et~al\mbox{.}}{2000}]%
        {de2000computational}
\bibfield{author}{\bibinfo{person}{Mark De~Berg}, \bibinfo{person}{Marc
  Van~Kreveld}, \bibinfo{person}{Mark Overmars}, {and}
  \bibinfo{person}{Otfried~Cheong Schwarzkopf}.}
  \bibinfo{year}{2000}\natexlab{}.
\newblock \showarticletitle{Computational geometry}.
\newblock In \bibinfo{booktitle}{{\em Computational geometry}}.
  \bibinfo{publisher}{Springer}, \bibinfo{pages}{1--17}.
\newblock


\bibitem[\protect\citeauthoryear{Isaacson}{Isaacson}{2005}]%
        {isaacson2005you}
\bibfield{author}{\bibinfo{person}{Eric~J Isaacson}.}
  \bibinfo{year}{2005}\natexlab{}.
\newblock \showarticletitle{What You See Is What You Get: on Visualizing
  Music.}. In \bibinfo{booktitle}{{\em ISMIR}}. Citeseer,
  \bibinfo{pages}{389--395}.
\newblock


\bibitem[\protect\citeauthoryear{Leskovec, Rajaraman, and Ullman}{Leskovec
  et~al\mbox{.}}{2014}]%
        {leskovec2014mining}
\bibfield{author}{\bibinfo{person}{Jure Leskovec}, \bibinfo{person}{Anand
  Rajaraman}, {and} \bibinfo{person}{Jeffrey~David Ullman}.}
  \bibinfo{year}{2014}\natexlab{}.
\newblock \bibinfo{booktitle}{{\em Mining of massive datasets}}.
\newblock \bibinfo{publisher}{Cambridge University Press}.
\newblock


\bibitem[\protect\citeauthoryear{{R. Cilibrasi P.M.B. Vitanyi}}{{R. Cilibrasi
  P.M.B. Vitanyi}}{2004}]%
        {Vitanyi2004}
\bibfield{author}{\bibinfo{person}{R~de~Wolf {R. Cilibrasi P.M.B. Vitanyi}}.}
  \bibinfo{year}{2004}\natexlab{}.
\newblock \showarticletitle{{Algorithmic clustering of music based on string
  compression}}.
\newblock \bibinfo{journal}{{\em Computer Music Journal\/}}
  \bibinfo{volume}{28}, \bibinfo{number}{4} (\bibinfo{year}{2004}),
  \bibinfo{pages}{49--67}.
\newblock
\showISSN{0148-9267}
\showDOI{%
\url{https://doi.org/10.1162/0148926042728449}}


\bibitem[\protect\citeauthoryear{Read}{Read}{1972}]%
        {read1972music}
\bibfield{author}{\bibinfo{person}{Gardner Read}.}
  \bibinfo{year}{1972}\natexlab{}.
\newblock \bibinfo{booktitle}{{\em Music notation: a manual of modern
  practice}}.
\newblock \bibinfo{publisher}{Rodale Press}.
\newblock


\bibitem[\protect\citeauthoryear{Rousseeuw}{Rousseeuw}{1987}]%
        {rousseeuw1987silhouettes}
\bibfield{author}{\bibinfo{person}{Peter~J Rousseeuw}.}
  \bibinfo{year}{1987}\natexlab{}.
\newblock \showarticletitle{Silhouettes: a graphical aid to the interpretation
  and validation of cluster analysis}.
\newblock \bibinfo{journal}{{\em Journal of computational and applied
  mathematics\/}}  \bibinfo{volume}{20} (\bibinfo{year}{1987}),
  \bibinfo{pages}{53--65}.
\newblock


\bibitem[\protect\citeauthoryear{Shi and Malik}{Shi and Malik}{2000}]%
        {shi2000normalized}
\bibfield{author}{\bibinfo{person}{Jianbo Shi} {and} \bibinfo{person}{Jitendra
  Malik}.} \bibinfo{year}{2000}\natexlab{}.
\newblock \showarticletitle{Normalized cuts and image segmentation}.
\newblock \bibinfo{journal}{{\em IEEE Transactions on pattern analysis and
  machine intelligence\/}} \bibinfo{volume}{22}, \bibinfo{number}{8}
  (\bibinfo{year}{2000}), \bibinfo{pages}{888--905}.
\newblock


\bibitem[\protect\citeauthoryear{Thickstun, Harchaoui, and Kakade}{Thickstun
  et~al\mbox{.}}{2017}]%
        {thickstun2017learning}
\bibfield{author}{\bibinfo{person}{John Thickstun}, \bibinfo{person}{Zaid
  Harchaoui}, {and} \bibinfo{person}{Sham Kakade}.}
  \bibinfo{year}{2017}\natexlab{}.
\newblock \showarticletitle{Learning Features of Music from Scratch}. In
  \bibinfo{booktitle}{{\em International Conference on Learning Representations
  (ICLR)}}.
\newblock


\end{thebibliography}
